\documentclass[aps,prl,twocolumn,superscriptaddress,preprintnumbers,nofootinbib,amsmath,amssymb]{revtex4-2}

\usepackage{graphicx}
\usepackage{mathtools}
\usepackage{bm}
\usepackage{hyperref}
\usepackage{enumitem}

\usepackage{xcolor}

\newcommand{\beq}{\begin{equation}}
\newcommand{\eeq}{\end{equation}}

\begin{document}

\title{
Isocurvature-Free QCD Axion Dark Matter from Inflaton-Driven Early QCD: \\ the Necessity of Inflationary Plateaus
}

\author{Katherine Freese}
\affiliation{Texas Center for Cosmology and Astroparticle Physics, Weinberg Institute for Theoretical Physics, Department of Physics, The University of Texas at Austin, Austin, TX, 78712, USA}
\affiliation{The Oskar Klein Centre, Department of Physics, Stockholm University, AlbaNova, SE-10691 Stockholm, Sweden}
\affiliation{Nordic Institute for Theoretical Physics (NORDITA), 106 91 Stockholm, Sweden}

\author{Evangelos I. Sfakianakis}
\affiliation{Texas Center for Cosmology and Astroparticle Physics, Weinberg Institute for Theoretical Physics, Department of Physics, The University of Texas at Austin, Austin, TX, 78712, USA}
\affiliation{Department of Physics, Harvard University, Cambridge, MA, 02131, USA}

\author{Barmak Shams Es Haghi}
\affiliation{Texas Center for Cosmology and Astroparticle Physics, Weinberg Institute for Theoretical Physics, Department of Physics, The University of Texas at Austin, Austin, TX, 78712, USA}
\affiliation{Department of Physics and Astronomy, University of Utah, Salt Lake City, UT, 84112, USA}

\begin{abstract}
A direct coupling between the inflaton and Standard Model gluons can dynamically raise the QCD confinement scale during inflation, making the axion temporarily heavy and suppressing axion isocurvature perturbations. As inflation proceeds, the confinement scale relaxes, the axion becomes light, and late-time de Sitter fluctuations can generate the observed dark matter abundance. We analyze this mechanism without specifying an inflationary potential, instead parametrizing the background by $\epsilon(N) \propto 1/N^p$, where $N$ is the number of $e$-folds before the end of inflation. The single parameter $p$ distinguishes  monomial models ($p=1$), standard plateau models ($p=2$), and ultra-flat plateau or hilltop-like models ($p\ge 3$). We analytically show  that the mechanism selects plateau-like ($p\ge 2$) inflation: monomial models generically cause the confinement scale to grow too rapidly, while plateau models keep the QCD sector under perturbative control. In the minimal scenario, reheating occurs through the same inflaton-gluon coupling, and viable axion dark matter production is obtained when deconfinement occurs after the CMB window. {The early-confinement sector also generically shifts the scalar spectral index to smaller, redder values. Because ultra-flat ($p \ge 3$) models inherently predict overly red spectra, this shift exacerbates their tension with CMB data, leaving $p=2$ plateau models  as the phenomenologically viable parameter space (in this parametrization).} 
\end{abstract}

\maketitle

\textit{\bf Introduction.}---%
The Peccei-Quinn (PQ) solution~\cite{Peccei:1977hh,Peccei:1977ur} to the strong CP problem introduces a global U(1)$_{\rm PQ}$ symmetry spontaneously broken at a high energy scale $f_a$. This yields the QCD axion~\cite{Weinberg:1977ma,Wilczek:1977pj,  Zhitnitsky:1980tq, Dine:1981rt, Preskill:1982cy,Abbott:1982af, Dine:1982ah}, an excellent cold dark matter (DM) candidate. 

The resulting cosmology depends critically on whether the PQ symmetry breaks before or after inflation~\cite{OHare:2024nmr}. In the pre-inflationary scenario, the axion field is homogenized, avoiding topological defects. However, as a light spectator field, it acquires quantum fluctuations of order $H / (2\pi)$ during inflation, where $H$ is the Hubble scale. These super-horizon modes follow a random walk, generating DM isocurvature perturbations. Satisfying the \textit{Planck} limit ($\beta_{\rm iso} < 0.038$) typically restricts the scale of inflation to $H \lesssim 10^8$ GeV, in  tension with high-scale inflationary models.

A mechanism to escape this bound is to increase the axion mass during inflation, exponentially suppressing its fluctuations. Since the axion mass scales with the QCD confinement scale, $m_a \propto \Lambda_{\rm QCD}^2 /f_a$, an enhanced axion mass can be achieved by making $\Lambda_{\rm QCD}$ dynamical in the early universe~\cite{Dvali:1995ce, Ipek:2018lhm, Dvali:2026ceb, Jeong:2013xta, Higaki:2014ooa}. 
The QCD sector is confined during the observable part (in the Cosmic Microwave Background) of inflation with large values of $m_a$ and $\Lambda_{\rm QCD}$; then deconfined at a later time in inflation and at the high temperatures after inflation;
then confined once the temperature of the universe drops to $T<\Lambda_0$, today's value of the QCD scale. 

In this Letter, we study  the inflaton itself as the driver of this dynamical confinement. We relate the field excursion directly to a generic parameterization of the slow-roll parameter 
$\epsilon(N) \propto 1/N^p$, where $N$ is the number of $e$-folds before the end of inflation. The  dimensionless parameter $p$ distinguishes  monomial models ($p=1$), standard plateau models ($p=2$), and ultra-flat plateau or hilltop-like models ($p\ge 3$).
We derive universal conditions required for the mechanism to succeed without spoiling the inflationary background. We demonstrate that this physical mechanism does not work with monomial inflation, whereas it perfectly accommodates standard plateau models.
{Further, we show that the predicted scalar spectral index ($n_s$) of inflaton density perturbations is shifted to lower, redder values. Because ultra-flat plateau models ($p \ge 3$) already predict overly red spectra, this shift exacerbates their tension with cosmic microwave background (CMB) data. We therefore demonstrate that standard $p=2$ plateau models are robustly selected as the best candidates for the success of this mechanism.}
We finally showcase the dependence of the various energy scales on the overall magnitude of the slow-roll parameter $\epsilon(N)$ for plateau and ultra-flat plateau models. An extended analysis will be presented  in Ref.~\cite{paper}.

During the completion of the current work (which we began in April 2025), Ref.~\cite{Dvali:2026ceb} appeared, using the same interaction term, but  within $SU(5)$ grand unified theory. In contrast, we take a phenomenological model-independent  approach. We therefore show (a) the necessity of plateau models for the success of this mechanism and (b) the resulting generic negative shift in $n_s$.


\medskip

\textit{\bf Inflaton-Driven Confinement.}---%
We introduce a non-renormalizable interaction between the inflaton $\phi$ and the Standard Model (SM) gluon field strength $G_{\mu\nu}$:
\begin{equation}
\label{eq:coupling}
  {\cal L}_g  \supset -{1\over 4}\left ({1\over g_{s0}^2} +{ \phi\over M} \right ) G_{\mu\nu}G^{\mu\nu} \, ,
\end{equation}
where $M$ is the mass scale of new physics.
The above inflaton-gluon coupling can arise for example from integrating out  heavy beyond SM fields, which will define the scale $M$, see e.g.~Ref.~\cite{Dvali:2026ceb}. 
During inflation, the classical homogeneous background $\bar \phi(N)$ changes the gluon kinetic term. This modifies the running of the strong coupling $\alpha_s$, yielding a confinement scale exponentially sensitive to the field value. We will take a negative field value for the inflaton.
Assuming the inflaton rolls toward the origin ($\bar \phi < 0$), the elevated confinement scale is~\cite{Ipek:2018lhm}
\begin{equation}
    \Lambda (\bar \phi) = \Lambda_0 \exp \left( \frac{24\pi^2}{2n_f-33} \frac{\bar \phi}{M} \right) \simeq \Lambda_0 e^{ 11 |\bar \phi|/M} \, ,
    \label{eq:LambdaQCD}
\end{equation}
where  $n_f=6$ is the number of active quarks and $\Lambda_0\simeq 400$ MeV is the standard QCD confinement scale at $\bar\phi = 0$. Because $\bar \phi$ modifies the confinement scale, the axion mass $m_a \sim \Lambda^2/f_a$ is correspondingly enhanced during inflation compared to the standard case without gluon coupling.
As the inflaton rolls down its potential, one can see that the value of $\Lambda$ decreases following the above equation.

In de Sitter space, the causal horizon is $R_H \sim H^{-1}$, corresponding to an ambient Gibbons-Hawking temperature $T_{\rm GH} = H/(2\pi)$~\cite{Gibbons:1977mu}. For QCD to confine, the characteristic size of a hadronic bound state ($\Lambda^{-1}$) must fit within the horizon; equivalently, the background temperature must be colder than the critical scale ($H \lesssim \Lambda$). If $\Lambda(\bar \phi) < H$, the spacetime expands too rapidly for quarks to causally communicate, melting the QCD vacuum and exponentially suppressing the instanton effects responsible for the axion mass and  rendering the axion effectively massless. 

Therefore, the axion can only acquire a heavy mass ($m_a > 3H/2$) capable of suppressing isocurvature perturbations while $\Lambda(\bar \phi) > H$. This is the case at the onset of inflation.  We define the deconfinement time, $N_{\rm dec}$ (in $e$-folds before the end of inflation), as the exact moment when $\Lambda(N_{\rm dec}) = H$. Hence, once the inflaton has rolled down the potential beyond the deconfinement time, i.e. for $N < N_{\rm dec}$, the axion  becomes an effectively massless spectator field and accumulates de Sitter fluctuations. The variance at the end of inflation is dominated by these late-time modes:
\beq
\label{eq:sigmatheta}
\sigma_\theta^2\equiv \langle \theta^2\rangle \simeq {H^2\over 4\pi^2 f_a^2} N_{\rm dec} \, .
\eeq
These fluctuations are innocuous as far as observable isocurvature modes are concerned, because they are on far too small scales to be seen in the CMB.
However, they are important as they
excite the coherent oscillations that act as cold DM, yielding an abundance
$\Omega_a\sim (f_a / 10^{12}\, {\rm GeV})^{7/6}\sigma_\theta^2$. Requiring that the axion makes up all the observed DM, $\Omega_a=\Omega_{DM} \sim 0.25$, and using Eq.~\eqref{eq:sigmatheta}, the axion decay constant  becomes
\beq
\label{eq:fa}
{f_a}\sim 10^7 \left( N_{\rm dec} {H^2\over M_{\rm Pl}^2}\right)^{6/5} M_{\rm Pl} \, ,
\eeq
 where $M_{\rm {Pl}} \simeq 2.4 \times 10^{18}$ GeV is the reduced Planck mass.
Larger $N_{\rm dec}$ for a given $H$ leads to  larger $f_a$.

Following inflation, the $\phi G_{\mu\nu}G^{\mu\nu}$ coupling provides a built-in reheating channel. The perturbative decay rate $\Gamma_{\phi\to gg} = m_\phi^3 / (8 \pi M^2)$ yields a standard radiation-dominated bath with $T_{{\rm reh}} \simeq 0.1 (m_\phi/M) \sqrt{m_\phi M_{\rm Pl}}$, where $m_\phi$ is the inflaton mass. In this minimal scenario, no beyond-SM fermions or auxiliary fields are introduced for purposes of reheating. The necessary coupling for increasing the QCD scale during inflation is also sufficient for reheating the universe after inflation. Further reheating channels will be presented in Ref.~\cite{paper}.


\medskip

\textit{\bf Hierarchy of Scales.}---%
For this framework to be theoretically consistent and successfully suppress axion isocurvature perturbations, the relevant mass and energy scales must satisfy a strict hierarchy during the CMB-observable epoch of inflation. We  define $\Lambda_{\rm CMB} \equiv \Lambda(N_{\rm CMB})$ as the QCD scale during the time during inflation where the CMB modes exited the horizon, typically for $N_{\rm CMB}\sim 50-60$.

First, the QCD sector must be confined early on to generate the axion potential, necessitating $\Lambda_{\rm CMB} > H$. However, its energy density must remain strictly subdominant to the inflationary background, $\Lambda_{\rm CMB}^4 \ll V$, to avoid dominating the universe's dynamics. Together, these bound the confinement scale: $H < \Lambda_{\rm CMB} \ll V^{1/4}$. 

Second, the axion must be sufficiently heavy to exponentially suppress super-horizon isocurvature fluctuations at CMB scales, requiring $m_{a,{\rm CMB}} > 3H/2$, where $m_{a,{\rm CMB}} \sim \Lambda_{\rm CMB}^2/f_a$. Perturbative unitarity of the axion cosine potential further demands $\Lambda_{\rm QCD} \lesssim f_a$.

Third, to prevent the restoration of the PQ symmetry (and the subsequent catastrophic production of topological defects) after inflation, the maximum temperature attained by the primordial plasma during reheating must remain below the PQ scale, $T_{\max} \lesssim f_a$. Finally, for the effective field theory describing the inflaton-gluon coupling to remain under theoretical control, the suppression scale must be sub-Planckian, $M \lesssim M_{\rm Pl}$.

We can summarize these necessary theoretical and phenomenological conditions into a single sequence of mass and energy scales during the CMB-observable epoch of inflation:
\begin{equation}
    3H/2 \lesssim m_{a,{\rm CMB}} < \Lambda_{\rm CMB} < \{f_a, V^{1/4}\} \, ,
    \label{eq:hierarchy}
\end{equation}
alongside $M \lesssim M_{\rm Pl}$ and $T_{\max} \lesssim f_a$, where $T_{\rm max}$ is the maximum temperature reached during reheating~\cite{Giudice:2000ex}. If this hierarchy is maintained, the theory is perturbative, the inflaton dominates the background dynamics, and the axion isocurvature perturbations are safely erased from the CMB.  Finally we note that at some point after the CMB-observable epoch, these conditions no longer hold; instead,  both $\Lambda<H\, ,~ m_a <3 H/2$ in order for axion DM to result from 
de Sitter excitations
of the axion field.


\medskip

\textit{\bf Universal Slow-Roll Constraints.}---%
Rather than assuming a specific potential $V(\phi)$, we parameterize the background using the slow-roll parameters $\epsilon(N)$ and $\eta(N)$. The field excursion is governed by $d\phi/dN =- M_{\rm Pl} \sqrt{2\epsilon(N)}$, where the field $\phi$ is taken to be negative during inflation and roll towards the origin, thus $d\phi/dN<0$.  Defining $\beta \equiv  11 M_{\rm Pl}/M$, the QCD energy density scales as $\Lambda^4 = \Lambda_0^4 e^{ - 4\beta \phi(N) / M_{\rm Pl}}$.

\begin{figure}[t]
    \centering
    \includegraphics[width=.96\linewidth]{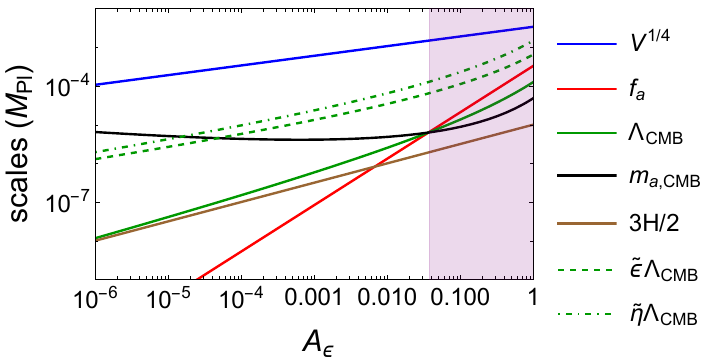}
    \\
    \includegraphics[width=\linewidth]{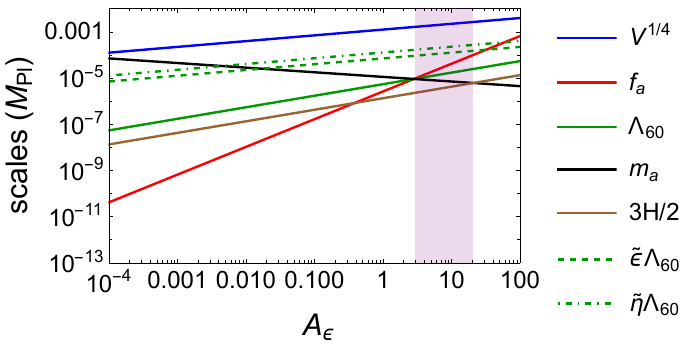}
    \caption{
    Relevant mass scales as a function of the 
    parameter $A_\epsilon $ that determines the inflationary Hubble scale (see Eqs.~\eqref{eq:epsilonparam},\eqref{eq:Hubblescale})  for $p=2$ (top), $p=3$ (bottom). The purple vertical band defines the successful parameter space: here  the sequence of scales given in Eq.~\eqref{eq:hierarchy} is satisfied, and the axion constitutes the entirety of DM as described above Eq~\eqref{eq:fa}.
For  $p=2$ (top) we used an $\alpha$-attractor model (which fixes $C=\sqrt{3\alpha/2} \ln( 8 M_{\rm Pl}^2 / 3\alpha )$ in Eq.~\eqref{eq:p2phi}; see Appendix) with $N_{\rm dec}=40$. 
{For $p = 3$ (bottom), we used $M/M_{\text{Pl}} = 0.5\sqrt{A_\epsilon}$ and $N_{\text{dec}} = 40$. With this choice of $M$, the QCD backreaction on the slow-roll parameters is  negligible. While the purple band demonstrates that ultra-flat models can successfully satisfy the hierarchy of  scales in Eq.~\eqref{eq:hierarchy}  to suppress isocurvature and produce axion dark matter, they remain observationally excluded due to their inherently red baseline spectral tilt ($n_s \simeq 0.950$).}
    }
    \label{fig:alphascan}
\end{figure}

The total effective potential for the inflaton plus QCD sector {during the confined phase \footnote{
To normalize correctly we should add in today's value $\Lambda_0 \sim (400 \, {\rm MeV})^4$, which is negligible compared to the high scales during inflation.
} 
is $V_{\rm tot}(\phi) = V(\phi) - \Lambda(\phi)^4$. The minus sign arises because the confined phase of QCD is a lower energy state than the deconfined (quark-gluon plasma) state~\cite{vonHarling:2017yew}.} For the mechanism to be valid, the QCD sector must not dominate the energy density ($\Lambda^4 \ll V$), nor can its derivatives spoil the slow-roll dynamics. The corrected parameters are
\begin{align}
    \epsilon_{\rm tot} &\simeq \left( \sqrt{\epsilon(N)} - 2\sqrt{2}\beta \frac{\Lambda(N)^4}{V(N)} \right)^2 \, , \label{eq:eps_tot} \\
    \eta_{\rm tot} &\simeq \eta(N) - 16\beta^2 \frac{\Lambda(N)^4}{V(N)} \, . \label{eq:eta_tot}
\end{align}
We require the correction terms to be strictly subdominant to the baseline parameters across the observable  $e$-folds of inflation. This imposes the 
backreaction 
constraints:
\begin{equation}
    \frac{\Lambda(N)^4}{V(N)} \ll \frac{\sqrt{\epsilon(N)}}{2\sqrt{2}\beta} \equiv {1\over \tilde \epsilon^4 } \, ,
~ \frac{\Lambda(N)^4}{V(N)} \ll \frac{|\eta(N)|}{16\beta^2} \equiv {1\over \tilde\eta^4}\, , 
\label{eq:backreaction_bounds}
\end{equation}
which will be written as $\tilde \epsilon \Lambda \ll V^{1/4}$ and $\tilde \eta \Lambda \ll V^{1/4}$, to be plotted in Figure~\ref{fig:alphascan}.

\medskip



\textit{\bf Parameterizing Inflationary Models.}---%
We evaluate the backreaction constraints by adopting a generic power-law parameterization\footnote{Such a parametrization is valid for ``single clock" models such as single-field rolling models of inflation.} for the first slow-roll parameter (see e.g.~\cite{Galante:2014ifa}) \beq 
\label{eq:epsilonparam}
\epsilon(N) = {A_\epsilon \over N^p} \, ,
\eeq   
where $A_\epsilon$ is a dimensionless constant.
Using 
\begin{equation}
\label{eq:fieldderivative}
\epsilon = {1\over 2} {\dot\phi^2 \over  M_{\rm Pl}^2H^2} ={1\over 2M_{\rm Pl}^2} \left ({d\phi\over dN}\right )^2 \, ,
\end{equation}
we can use the functional form of $\epsilon(N)$ to derive the inflaton evolution $\phi(N)$.\footnote{Since the derived $\phi(N)$ assumed a slow-roll solution $\epsilon\propto N^p$, which fails
for $N = 0$, the  expression for $\phi(N)$ cannot be used
for $\phi\approx  0$. An ad-hoc solution would be to substitute
$N \to N + N_0$, which makes $\phi(N = 0)$ well behaved. This
detail is not important for our discussion, which contains freedom in the form of integration constants, that are determined only by choosing a specific inflationary potential.}

Crucially, the parameter $A_\epsilon$ is not merely a mathematical construct; it acts as a master key that strictly dictates the entire set of inflationary observables. In single-field inflation, the amplitude of the primordial scalar curvature power spectrum is given by $A_s = H_*^2 / (8\pi^2 M_{\rm Pl}^2 \epsilon_*)$, evaluated when the CMB pivot scale exits the horizon at $N_* \simeq 60$. Because $A_s \simeq 2.1 \times 10^{-9}$ is fixed by precise \textit{Planck} measurements, the inflationary Hubble scale $H_*$ is rigidly locked to the value of the slow-roll parameter at that epoch. 

By substituting our parameterization $\epsilon_* = A_\epsilon / N_*^p$, we can invert this relationship to express the inflationary energy scale directly as:
\begin{equation}
\label{eq:Hubblescale}
    H_* \simeq \pi M_{\rm Pl} \sqrt{8 A_s \frac{A_\epsilon }{N_*^p}} \, .
\end{equation}
Because plateau models ($p \ge 2$) are exceptionally flat, $H$ remains nearly constant over the observable $e$-folds, allowing us to approximate $H \simeq H_*$ throughout.

Thus, $A_\epsilon $ defines the inflationary Hubble scale, for a given value of $N_*$. Furthermore, $A_\epsilon $ directly determines the tensor-to-scalar ratio for single field models with unit sound speed, $r = 16A_\epsilon  / N_*^p$. For the case of plateau models ($p \geq 2$),
where $\epsilon \ll |\eta|$, the slow-roll consistency relation $\eta = 2\epsilon + (2\epsilon)^{-1} d\epsilon/dN$ 
uniquely fixes the second slow-roll parameter to $\eta(N) \simeq -p / (2N)<0$, where we dropped terms ${\cal O}(1/N^p)$ in $\eta$. This completely determines the scalar spectral tilt, $n_s \simeq 1 + 2\eta$.

By selecting $p$ and $A_\epsilon $, the entire cosmological background—$H$, $r$, and $n_s$—is fully specified. The QCD scale $\Lambda(N)$ inherits an integration constant $C$ from the field trajectory $\phi(N)$ as we will show later. However, once a specific inflationary model is chosen, $C$ is fixed.
This leaves only the coupling mass scale $M$ to dictate the exact timing of the QCD deconfinement transition.
Overall, three parameters define the dynamics: $p$, $A_\epsilon$ and $M$, up to a possible integration constant in $\phi(N)$.

\medskip

\textbf{\bf Monomial Models ($p < 2$):}
For models with a monomial potential $V(\phi)\propto \phi^n$, the field excursion is  integrated to give $|\phi(N)|\simeq \sqrt{2nN}M_{\rm Pl}$ and the first slow roll parameter $\epsilon \simeq n/(4N)$. This corresponds to $p=1$ in our parametrization of Eq.~\eqref{eq:epsilonparam}, where $\epsilon\propto 1/N$ and $\phi(N)\propto \sqrt{N}$.
 Inserted into Eq.~\eqref{eq:LambdaQCD}, the QCD energy density $\Lambda(N)^4$ grows \textit{exponentially} with $N$ (where, again, $N$ counts the number of $e$-folds back from the end of inflation defined as $N=0$). Because the inflationary potential $V(N)$ only grows polynomially $V\propto \phi^n\propto N^{n/2}$, the QCD sector rapidly overtakes the inflaton energy density as we look back up the inflationary potential to earlier times. The fundamental zeroth-order requirement $\Lambda^4 \ll V$ is violently violated deep in inflation unless the coupling scale is tuned to super-Planckian values, $M \gg M_{\rm Pl}$. However, if $M$ is made large enough to suppress the exponential growth, the confinement scale never rises significantly above its vacuum value $\Lambda_0 \sim 400$ MeV. Consequently, $\Lambda \ll H$ during the CMB epoch, the axion remains light, and the isocurvature perturbations are entirely unsuppressed.
We have thus shown that this mechanism is inherently incompatible with monomial inflation.

\medskip

\textbf{Standard Plateau Models ($p = 2$):}
For plateau models (e.g., $\alpha$-attractors~\cite{Carrasco:2015uma, Carrasco:2015pla,Galante:2014ifa, Kallosh:2013yoa, Kallosh:2013hoa, Iarygina:2018kee}, Higgs inflation~\cite{Bezrukov:2007ep, Rubio:2018ogq, Greenwood:2012aj} and non-minimally coupled models~\cite{Kaiser:2013sna, Christodoulidis:2025wew}), $p=2$ yields $\eta \simeq -1/N$ in the standard QCD case (without the inflaton/gluon coupling), predicting $n_s \simeq 1 + 2\eta \simeq 0.967$ at $N_*=60$, in excellent agreement with \textit{Planck}. 

\begin{figure}[t]
    \centering
    \includegraphics[width=.8\linewidth]{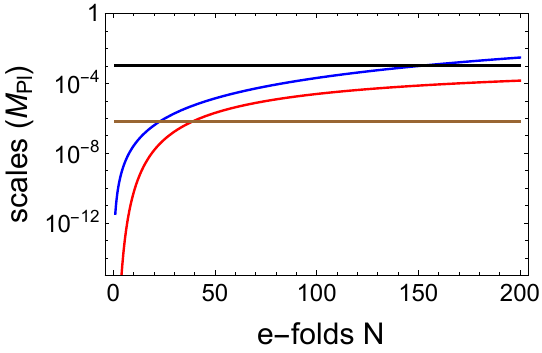}
    \caption{Quantities of interest as a function of $N$, the number of $e$-folds before the end of inflation, for a specific choice of parameters described below.
    The QCD confinement scale $\Lambda(N)$ for $p=2,3$ is shown in blue and red curves respectively, where the first slow-roll parameter is characterized by the power-law index $p$ as $\epsilon(N) = A_\epsilon /N^p$. The brown and black curves show the  values of $H$ and $V^{1/4}$ respectively, per Eq.~\eqref{eq:Hubblescale}. 
   We chose $M=0.4 M_{\rm Pl}$ and $A_\epsilon=0.01$ for $p=2$, leading to $H\sim 6\times 10^{-7}M_{\rm Pl} $ for $N_*=60$. To keep the same Hubble scale, we chose $A_\epsilon=0.6$ for $p=3$. While the overall amplitude of the blue and red curves depends on the integration constant of $\phi(N)$ in Eqs.~\eqref{eq:p2phi},\eqref{eq:p3phi}, their different growth rate, monomial and bounded respectively, is as shown and only depends on $p$.  Importantly, for both $p=2,3$, the QCD sector does not dominate the energy density ($\Lambda^4 \ll V$) during the last $60$ $e$-folds of inflation. Pictorially, $N_{\rm dec}$ is the time when the blue/red curves corresponding to $\Lambda(N)$ intersect the horizontal brown line, corresponding to $H$. 
    }
    \label{fig:scales}
\end{figure}

In all cases (with our without inflaton/gluon coupling), integrating the field excursion in Eqn. \eqref{eq:fieldderivative} yields
\beq
\label{eq:p2phi}
\phi(N) = -M_{\rm Pl} \sqrt{2A_\epsilon } \ln N - C \, ,
\eeq where $C$ is an integration constant, which can depend on $A_\epsilon $ (the exact relation for $\alpha$-attractors is given in the Appendix).  
The exponential mapping of the inflaton-gluon coupling perfectly cancels the logarithmic dependence of $\phi$ on N, leading to
\begin{equation}
\label{eq:Lambdap2}
    \Lambda(N) = \tilde{\Lambda}_0 N^\gamma \, , \quad \gamma = \frac{11 M_{\rm Pl}}{M} \sqrt{2A_\epsilon } \, , 
\end{equation}
where $\tilde{\Lambda}_0 \equiv \Lambda_0 e^{11 C / M}$. The QCD scale experiences purely benign power-law growth,  as shown in Figure~\ref{fig:scales}). This  allows the model to satisfy the backreaction constraints in Eq.~\eqref{eq:backreaction_bounds}, $\Lambda^4/V \ll {\rm Min}\left [
1, \sqrt{\epsilon}/(2\sqrt 2\beta) , |\eta|/(16\beta^2)
\right ]$, during the  observable period of inflation.

Crucially, the integration constant $C$ represents the exact field offset from the minimum, which varies between specific models, which otherwise give the same $\epsilon(N)$ behavior (see e.g.~Ref.~\cite{Iarygina:2020dwe} for the difference in $\phi$ at the end of inflation between the E and T model of $\alpha$-attractors). While $\tilde{\Lambda}_0$ is exponentially sensitive to $C$, this does not induce fine-tuning. The exact deconfinement condition $\Lambda(N_{\rm dec}) = H$ yields $M \propto [11 C + 11 M_{\rm Pl}\sqrt{2A_\epsilon }\ln N_{\rm dec}] / \ln(H/\Lambda_0)$. Because the denominator $\ln(H/\Lambda_0) \sim 30$ is logarithmically large, $\mathcal{O}(1)$ shifts in $C$ (in Planck units) only require $\mathcal{O}(10\%)$ adjustments to the free parameter $M$ to recover the exact same dynamics. 

Consequently, while the precise value of $M$ is model-dependent, the deconfinement time $N_{\rm dec}$ is universally robust and physically dictates the DM abundance. We therefore swap $M$ for $N_{\rm dec}$ as our primary phenomenological parameter. For $N_{\rm dec} \sim 40-45$ (corresponding to  gluon reheating), the $p=2$ framework perfectly suppresses isocurvature and reproduces the correct DM relic density across all standard plateau models for certain values of $ A_\epsilon $. It is worth noting that an upper bound can be invoked by demanding  $M<M_{\rm Pl}$, which we don't implement, as it depends sensitively on the integration constant $C$ and can thus be different for different members of the $p=2$ family. In Ref.~\cite{paper} we focus on the T model of $\alpha$-attractors and explicitly implement the cutoff $M<M_{\rm Pl}$, showing the viability of our general analytic formulas presented here for a concrete realization.

\medskip

\textbf{Ultra-Flat Plateau Models ($p \ge 3$):}
For potentials that become ultra-flat, deep into inflation (e.g., Quartic Hilltop models\footnote{Notice that we need to shift the usual quartic hilltop potential, such that $\phi<0$ during inflation and $\phi\simeq 0$ at the end of inflation or shift the coupling term in Eq~\eqref{eq:coupling} as ${\cal L }\subset -{1\over 4}(\phi- \phi_{\rm end}$)GG/M.} with $V(\phi) \propto (1-\phi^4/\mu^4)$ 
or supergravity frameworks with  kinetic poles~\cite{Terada:2016nqg,Roest:2013fha,Galante:2014ifa}), Eq.~\eqref{eq:epsilonparam} with $p=3$ applies. The field integrates to an inverse-root profile, \beq
\label{eq:p3phi}
\phi(N) = \phi_\infty + 2 M_{\rm Pl} \sqrt{2A_\epsilon \over N+N_0}  \, ,
\eeq
where $\phi_\infty<0$ corresponds to the limiting field value for an infinite number of $e$-folds of inflation and  $N_0$ was inserted such that $\phi(N=0)$ is finite.
The ``ultra-flatness" of the potential is evident by the fact that a finite field displacement leads to infinite $e$-folds of inflation. 
We should stress again, that our $\phi(N)$ is not an exact solution, since the parametrization $\epsilon\propto N^{-3}$ fails close to the end of inflation, but holds during the slow-roll phase.
Substituting $\phi(N)$ into the QCD  scale gives:
\begin{equation}
    \Lambda(N) = \Lambda_{\max} \exp\left( - \frac{\kappa}{\sqrt{N+N_0}} \right) \, ,
\end{equation} 
where $\Lambda_{\max} \equiv \Lambda_0 \exp(11 \phi_\infty/M)$ and $\kappa \equiv 22 M_{\rm Pl} \sqrt{2A_\epsilon } / M$. Because the potential is exceptionally flat, the total field excursion is strictly bounded by the asymptotic value $\phi_\infty$. Consequently, deep into inflation ($N \to \infty$), the confinement scale approaches a hard, constant upper limit $\Lambda_{\max}$. This provides absolute analytical protection against exponential backreaction, allowing the slow-roll dynamics to remain unspoiled, even infinitely deep in the plateau (see Figure~\ref{fig:scales}).

\medskip

\textit{\bf Shifting the Spectral Tilt.}---%
In the standard case without coupling to gluons, {the two main families of models $p=2,3$ provide sharp predictions for the scalar spectral index, $n_s \simeq 1 - p/N_* $, leading to $n_s\simeq0.967, 0.95$ and for  $p=2,3$ respectively, both evaluated at $N_*=60$. We see that the $n_s|_{p=3}$  is overly red and in tension with \textit{Planck} measurements.
Interestingly, the coupling to gluons proposed in this paper, due to the
negative sign of the QCD backreaction, only exacerbates this problem. 
From Eq.~\eqref{eq:eta_tot}, the second derivative of the exponentially growing confinement scale yields a strictly negative contribution to $\eta_{\rm tot}$. Consequently, the QCD sector inevitably shifts the scalar spectral tilt, $n_s \simeq 1 + 2\eta_{\rm tot}$, to smaller redder values.
}
\\
{This shift can be constrained to $\lesssim \mathcal{O}(10^{-2})$ by keeping the ratio $\Lambda^4 / V$ sufficiently small. For standard $p=2$ models in the case without inflaton/gluon coupling, which typically predict a spectral index of $n_s \simeq 0.967$ (at $N=60$), a small backreaction is in principle safe. Furthermore, depending on the coupling scale $M$, this strict negative shift provides a distinct observational signature.
On the other hand, $p \ge 3$ models inherently suffer from an overly red spectral tilt ($n_s \simeq 0.950$). Because the QCD backreaction can only yield a negative shift, this mechanism  leaves them in tension (or disagreement) with current data. Therefore, while $p \ge 3$ models elegantly bound the exponential growth of $\Lambda$, they cannot be reconciled with CMB observations. This phenomenological signature restricts the viability of this early-confinement mechanism to $p=2$ plateau models (given the parametrization in Eq. \eqref{eq:epsilonparam}).}

\medskip

\textit{\bf A Successful Inflationary Realization.}---A successful model must satisfy the following requirements:
(i) suppression of isocurvature perturbations without spoiling slow-roll inflation requires the hierarchy of scales given in Eq.~\eqref{eq:hierarchy};  (ii) the axion must constitute the entirety of the DM (see Eq.~\eqref{eq:fa}) and (iii) $f_a >T_{\rm max}$.

In Figure~\ref{fig:alphascan}, we show examples of such models for $p=2$ (upper panel) and $p=3$ (lower panel).  The figure plots the relevant scales as a function of the parameter $A_{\epsilon}$ which dictates the inflationary scale as per Eqns.~\eqref{eq:epsilonparam},\eqref{eq:Hubblescale}.  The purple band is the region that satisfies the requirements (i), (ii) just described.\footnote{{Satisfying $f_a > T_{\text{max}}$ places a non-trivial constraint on the parameter space. A massive inflaton decaying to gluons leads to $T_{\text{max}} \sim (\Gamma_\phi M_{\text{Pl}}^2 H_{\text{end}})^{1/4} \sim m_\phi (M_{\text{Pl}}/M)^{1/2}$. For standard plateau models, 
$m_\phi \sim   10^{-6} M_{\text{Pl}}$. Because $M \lesssim M_{\text{Pl}}$, the bath inevitably spikes to $T_{\text{max}} \sim 10^{-5} M_{\text{Pl}}$. Therefore, avoiding thermal PQ restoration   requires $f_a \gtrsim 10^{-5} M_{\text{Pl}}$. Remarkably, as shown in Fig.~\ref{fig:alphascan}, this seamlessly aligns with the exact $f_a$ window dictated by the dark matter relic abundance in our mechanism.}}  In both upper and lower panels, we have chosen $N_{\rm dec}$ = 40, where $N_{\rm dec}$ is the deconfinement time in
$e$-folds before the end of inflation.
In general, in order to satisfy the hierarchy of Eq.~\eqref{eq:hierarchy}, a large value of $f_a$ is needed, which in turn prefers a large value of $N_{\rm dec}$ per Eq.~\eqref{eq:fa}. Thus, deconfinement must occur shortly after CMB scales leave the horizon.

In the upper panel of Figure~\ref{fig:alphascan}, as a specific realization of the $p=2$ family of models,
we use the T-model of $\alpha$-attractors. This fixes the integration constant in the inflaton trajectory to $C=\sqrt{3\alpha/2} \ln( 8 M_{\rm Pl}^2 / 3\alpha )$, as shown in the  Appendix,  leaving $M$ and $A_\epsilon$ as free parameters. Our choice of  $N_{\rm dec}$ then defines $M$ through solving Eq.~\eqref{eq:Lambdap2} for $\Lambda(N_{\rm dec})=H$. 
By using a concrete $\alpha$-attractor model, we also verified that $f_a>T_{\rm max}$. The purple region on Figure~\ref{fig:alphascan} shows the values of $A_\epsilon$ where all relevant scales follow Eq.~\eqref{eq:hierarchy}, thus the model successfully suppresses CMB isocurvature, while producing the observed DM. The allowed parameter space corresponds to 
 an
 axion decay constant $f_a \sim 10^{13}-10^{14}$ GeV
and an
inflationary Hubble scale $H \sim 10^{11}-10^{12}$ GeV. A detailed analysis of our mechanism in $\alpha$-attractors, including the effect of further reheating channels, will be shown  in Ref.~\cite{paper}.

{For the $p=3$ case (lower panel), we have chosen $M/M_{\text{Pl}} = 0.5\sqrt{A_\epsilon}$ such that the QCD backreaction remains negligible (in principle it may be larger). As shown by the purple band, the bounded nature of the $p \ge 3$ field excursion allows these models to easily satisfy the hierarchy of scales without spoiling the inflationary background. However, because this minimal backreaction cannot cure (and a larger backreaction would only exacerbate) their inherently red spectral index ($n_s \simeq 0.950$), this theoretically viable parameter space remains observationally excluded. 
\\
We note that realistic realizations of $p=3$ inflation, such as the quartic hilltop model, require higher-order terms to bound the potential from below and allow for reheating. While it is known that specific tunings of these stabilizing terms can shift $n_s$ back into the observable window, doing so breaks the generic $p=3$ universality, introducing severe model-dependence. Thus, while one might engineer a completed hilltop potential to satisfy both CMB data and our confinement mechanism, the standard $p=2$ plateau models effortlessly accommodate all observational and theoretical constraints without any such tuning. This  reinforces why our mechanism naturally isolates $p=2$ models as the most elegant realizations.
}


\bigskip

\textit{\bf Conclusion.}---%
We have presented a model-independent analysis of an inflaton-driven early QCD confinement mechanism designed to solve the axion isocurvature problem. We proved analytically that this framework inherently selects plateau-like inflationary models ($\epsilon \propto 1/N^p$ with $p \ge 2$). Models with a monomial potential ($p<2$) fail because the field excursion causes the confinement scale to grow exponentially with $N$, rapidly violating slow-roll constraints.  Standard plateau models ($p=2$) yield a benign polynomial growth of the QCD scale, while steep plateau models ($p \ge 3$) mathematically bound the confinement scale from above. Thus for both $p=2,3$, the QCD sector does not dominate the energy density ($\Lambda^4 \ll V$) during the last $60$ $e$-folds of inflation,  thereby allowing for successful inflation. Further, the QCD backreaction  yields a negative correction to $\eta$, shifting the scalar spectral tilt $n_s$ to smaller (redder) values. Consequently, ultra-flat plateau models ($p \ge 3$), which already predict overly red spectra, are pushed further out of alignment with CMB data. Therefore, for the parametrization of Eq. \eqref{eq:epsilonparam}, the combination of backreaction constraints and CMB observables uniquely isolates standard plateau models ($p=2$) as the viable family of models for this mechanism.
We note that our analysis is valid for models with a ``single clock," e.g. single field inflation models.  More complicated inflationary constructions e.g. with
multiple fields are possible and are not encompassed by the parametrization in Eq. \eqref{eq:epsilonparam}. A separate analysis would be required in such  models.

In its most economical setup, with purely gluon-driven reheating, the mechanism elegantly succeeds if deconfinement occurs roughly $40-45$ $e$-folds before the end of inflation. This provides an exciting avenue for future work, since modes that exit the horizon close to the deconfinement transition may appear in large multipoles on the CMB, depending on the exact post-inflationary expansion history of the universe. This can be a target for comparison with future data on large CMB multipoles from terrestrial experiments. A dedicated analysis, in addition to a study of other reheating channels, will be presented in upcoming papers.

\vspace{1em}
\begin{acknowledgments}
We thank Miguel Barroso~Varela for comments on the manuscript. K.F. holds the Jeff \& Gail Kodosky Endowed Chair at the University of Texas, Austin and is grateful for support. K.F. and E.I.S. acknowledge support from the U.S. Department of Energy, Office of Science, Office of High Energy Physics program under Award Number DESC-0022021.  B.S.E. is particularly grateful to Pearl Sandick, Prof of Physics and Dean of the College of Science at the University of Utah, for her continuous support and
encouragement.
\end{acknowledgments}

\bibliographystyle{apsrev4-2}
\bibliography{draft.bib}

@misc{Dvali:1995ce,
	archiveprefix = {arXiv},
	author = {Dvali, G. R.},
	eprint = {hep-ph/9505253},
	month = {5},
	reportnumber = {IFUP-TH-21-95},
	title = {{Removing the cosmological bound on the axion scale}},
	year = {1995}}

@article{Ipek:2018lhm,
	archiveprefix = {arXiv},
	author = {Ipek, Seyda and Tait, Tim M. P.},
	doi = {10.1103/PhysRevLett.122.112001},
	eprint = {1811.00559},
	journal = {Phys. Rev. Lett.},
	number = {11},
	pages = {112001},
	primaryclass = {hep-ph},
	reportnumber = {UCI-HEP-TR-2018-15},
	title = {{Early Cosmological Period of QCD Confinement}},
	volume = {122},
	year = {2019}}

@article{Iarygina:2018kee,
	archiveprefix = {arXiv},
	author = {Iarygina, Oksana and Sfakianakis, Evangelos I. and Wang, Dong-Gang and Achucarro, Ana},
	doi = {10.1088/1475-7516/2019/06/027},
	eprint = {1810.02804},
	journal = {JCAP},
	pages = {027},
	primaryclass = {astro-ph.CO},
	reportnumber = {Nikhef 2018-050},
	title = {{Universality and scaling in multi-field $\alpha$-attractor preheating}},
	volume = {06},
	year = {2019}}

@article{Carrasco:2015uma,
	archiveprefix = {arXiv},
	author = {Carrasco, John Joseph M. and Kallosh, Renata and Linde, Andrei and Roest, Diederik},
	doi = {10.1103/PhysRevD.92.041301},
	eprint = {1504.05557},
	journal = {Phys. Rev. D},
	number = {4},
	pages = {041301},
	primaryclass = {hep-th},
	title = {{Hyperbolic geometry of cosmological attractors}},
	volume = {92},
	year = {2015}}

@article{Carrasco:2015pla,
	archiveprefix = {arXiv},
	author = {Carrasco, John Joseph M. and Kallosh, Renata and Linde, Andrei},
	doi = {10.1007/JHEP10(2015)147},
	eprint = {1506.01708},
	journal = {JHEP},
	pages = {147},
	primaryclass = {hep-th},
	title = {{$\alpha $-Attractors: Planck, LHC and Dark Energy}},
	volume = {10},
	year = {2015}}

@article{Kallosh:2013yoa,
	archiveprefix = {arXiv},
	author = {Kallosh, Renata and Linde, Andrei and Roest, Diederik},
	doi = {10.1007/JHEP11(2013)198},
	eprint = {1311.0472},
	journal = {JHEP},
	pages = {198},
	primaryclass = {hep-th},
	title = {{Superconformal Inflationary $\alpha$-Attractors}},
	volume = {11},
	year = {2013}}

@article{Kallosh:2013hoa,
	archiveprefix = {arXiv},
	author = {Kallosh, Renata and Linde, Andrei},
	doi = {10.1088/1475-7516/2013/07/002},
	eprint = {1306.5220},
	journal = {JCAP},
	pages = {002},
	primaryclass = {hep-th},
	title = {{Universality Class in Conformal Inflation}},
	volume = {07},
	year = {2013}}

@article{Peccei:1977hh,
	author = {Peccei, R. D. and Quinn, Helen R.},
	doi = {10.1103/PhysRevLett.38.1440},
	journal = {Phys. Rev. Lett.},
	pages = {1440--1443},
	reportnumber = {ITP-568-STANFORD},
	title = {{CP Conservation in the Presence of Instantons}},
	volume = {38},
	year = {1977}}

@article{Peccei:1977ur,
	author = {Peccei, R. D. and Quinn, Helen R.},
	doi = {10.1103/PhysRevD.16.1791},
	journal = {Phys. Rev. D},
	pages = {1791--1797},
	reportnumber = {ITP-572-STANFORD},
	title = {{Constraints Imposed by CP Conservation in the Presence of Instantons}},
	volume = {16},
	year = {1977}}

@article{Weinberg:1977ma,
	author = {Weinberg, Steven},
	doi = {10.1103/PhysRevLett.40.223},
	journal = {Phys. Rev. Lett.},
	pages = {223--226},
	reportnumber = {HUTP-77/A074},
	title = {{A New Light Boson?}},
	volume = {40},
	year = {1978}}

@article{Wilczek:1977pj,
	author = {Wilczek, Frank},
	doi = {10.1103/PhysRevLett.40.279},
	journal = {Phys. Rev. Lett.},
	pages = {279--282},
	reportnumber = {Print-77-0939 (COLUMBIA)},
	title = {{Problem of Strong $P$ and $T$ Invariance in the Presence of Instantons}},
	volume = {40},
	year = {1978}}

@article{Preskill:1982cy,
	author = {Preskill, John and Wise, Mark B. and Wilczek, Frank},
	doi = {10.1016/0370-2693(83)90637-8},
	editor = {Srednicki, M. A.},
	journal = {Phys. Lett. B},
	pages = {127--132},
	reportnumber = {HUTP-82-A048, NSF-ITP-82-103},
	title = {{Cosmology of the Invisible Axion}},
	volume = {120},
	year = {1983}}

@article{Abbott:1982af,
	author = {Abbott, L. F. and Sikivie, P.},
	doi = {10.1016/0370-2693(83)90638-X},
	editor = {Srednicki, M. A.},
	journal = {Phys. Lett. B},
	pages = {133--136},
	reportnumber = {PRINT-82-0695 (BRANDEIS)},
	title = {{A Cosmological Bound on the Invisible Axion}},
	volume = {120},
	year = {1983}}

@article{Dine:1982ah,
	author = {Dine, Michael and Fischler, Willy},
	doi = {10.1016/0370-2693(83)90639-1},
	editor = {Srednicki, M. A.},
	journal = {Phys. Lett. B},
	pages = {137--141},
	reportnumber = {UPR-0201T},
	title = {{The Not So Harmless Axion}},
	volume = {120},
	year = {1983}}

@article{OHare:2024nmr,
	archiveprefix = {arXiv},
	author = {O'Hare, Ciaran A. J.},
	doi = {10.22323/1.454.0040},
	eprint = {2403.17697},
	journal = {PoS},
	pages = {040},
	primaryclass = {hep-ph},
	title = {{Cosmology of axion dark matter}},
	volume = {COSMICWISPers},
	year = {2024}}

@article{Giudice:2000ex,
	archiveprefix = {arXiv},
	author = {Giudice, Gian Francesco and Kolb, Edward W. and Riotto, Antonio},
	doi = {10.1103/PhysRevD.64.023508},
	eprint = {hep-ph/0005123},
	journal = {Phys. Rev. D},
	pages = {023508},
	reportnumber = {SNS-PH-00-05, FERMILAB-PUB-00-075-A, CERN-TH-2000-107},
	title = {{Largest temperature of the radiation era and its cosmological implications}},
	volume = {64},
	year = {2001}}

@misc{Dvali:2026ceb,
	archiveprefix = {arXiv},
	author = {Dvali, Gia and Fitz, Sophia and Komisel, Lucy},
	eprint = {2603.28620},
	month = {3},
	primaryclass = {hep-ph},
	title = {{Removing the Cosmological Bound on the Axion Scale via Confinement During Inflation}},
	year = {2026}}

@misc{Iarygina:2020dwe,
	archiveprefix = {arXiv},
	author = {Iarygina, Oksana and Sfakianakis, Evangelos I. and Wang, Dong-Gang and Ach{\'u}carro, Ana},
	eprint = {2005.00528},
	month = {5},
	primaryclass = {astro-ph.CO},
	title = {{Multi-field inflation and preheating in asymmetric $\alpha$-attractors}},
	year = {2020}}

@article{Gibbons:1977mu,
	author = {Gibbons, G. W. and Hawking, S. W.},
	doi = {10.1103/PhysRevD.15.2738},
	journal = {Phys. Rev. D},
	pages = {2738--2751},
	title = {{Cosmological Event Horizons, Thermodynamics, and Particle Creation}},
	volume = {15},
	year = {1977}}

@misc{Christodoulidis:2025wew,
	archiveprefix = {arXiv},
	author = {Christodoulidis, Perseas and Rosati, Robert and Sfakianakis, Evangelos I.},
	eprint = {2504.12406},
	month = {4},
	primaryclass = {astro-ph.CO},
	title = {{Robust non-minimal attractors in many-field inflation}},
	year = {2025}}

@article{Rubio:2018ogq,
	archiveprefix = {arXiv},
	author = {Rubio, Javier},
	doi = {10.3389/fspas.2018.00050},
	eprint = {1807.02376},
	journal = {Front. Astron. Space Sci.},
	pages = {50},
	primaryclass = {hep-ph},
	title = {{Higgs inflation}},
	volume = {5},
	year = {2019}}

@article{Bezrukov:2007ep,
	archiveprefix = {arXiv},
	author = {Bezrukov, Fedor L. and Shaposhnikov, Mikhail},
	doi = {10.1016/j.physletb.2007.11.072},
	eprint = {0710.3755},
	journal = {Phys. Lett. B},
	pages = {703--706},
	primaryclass = {hep-th},
	title = {{The Standard Model Higgs boson as the inflaton}},
	volume = {659},
	year = {2008}}

@article{Greenwood:2012aj,
	archiveprefix = {arXiv},
	author = {Greenwood, Ross N. and Kaiser, David I. and Sfakianakis, Evangelos I.},
	doi = {10.1103/PhysRevD.87.064021},
	eprint = {1210.8190},
	journal = {Phys. Rev. D},
	pages = {064021},
	primaryclass = {hep-ph},
	reportnumber = {PREPRINT-MIT-CTP-4411},
	title = {{Multifield Dynamics of Higgs Inflation}},
	volume = {87},
	year = {2013}}

@article{Kaiser:2013sna,
	archiveprefix = {arXiv},
	author = {Kaiser, David I. and Sfakianakis, Evangelos I.},
	doi = {10.1103/PhysRevLett.112.011302},
	eprint = {1304.0363},
	journal = {Phys. Rev. Lett.},
	number = {1},
	pages = {011302},
	primaryclass = {astro-ph.CO},
	reportnumber = {PREPRINT-MIT-CTP-4451},
	title = {{Multifield Inflation after Planck: The Case for Nonminimal Couplings}},
	volume = {112},
	year = {2014}}

@article{German:2020rpn,
	archiveprefix = {arXiv},
	author = {German, Gabriel},
	doi = {10.1088/1475-7516/2021/02/034},
	eprint = {2011.12804},
	journal = {JCAP},
	pages = {034},
	primaryclass = {astro-ph.CO},
	title = {{Quartic hilltop inflation revisited}},
	volume = {02},
	year = {2021}}

@article{Dimopoulos:2020kol,
	archiveprefix = {arXiv},
	author = {Dimopoulos, Konstantinos},
	doi = {10.1016/j.physletb.2020.135688},
	eprint = {2006.06029},
	journal = {Phys. Lett. B},
	pages = {135688},
	primaryclass = {hep-ph},
	title = {{An analytic treatment of quartic hilltop inflation}},
	volume = {809},
	year = {2020}}

@article{Roest:2013fha,
	archiveprefix = {arXiv},
	author = {Roest, Diederik},
	doi = {10.1088/1475-7516/2014/01/007},
	eprint = {1309.1285},
	journal = {JCAP},
	pages = {007},
	primaryclass = {hep-th},
	title = {{Universality classes of inflation}},
	volume = {01},
	year = {2014}}

@article{Adshead:2010mc,
	archiveprefix = {arXiv},
	author = {Adshead, Peter and Easther, Richard and Pritchard, Jonathan and Loeb, Abraham},
	doi = {10.1088/1475-7516/2011/02/021},
	eprint = {1007.3748},
	journal = {JCAP},
	pages = {021},
	primaryclass = {astro-ph.CO},
	title = {{Inflation and the Scale Dependent Spectral Index: Prospects and Strategies}},
	volume = {02},
	year = {2011}}

@article{Galante:2014ifa,
	archiveprefix = {arXiv},
	author = {Galante, Mario and Kallosh, Renata and Linde, Andrei and Roest, Diederik},
	doi = {10.1103/PhysRevLett.114.141302},
	eprint = {1412.3797},
	journal = {Phys. Rev. Lett.},
	number = {14},
	pages = {141302},
	primaryclass = {hep-th},
	title = {{Unity of Cosmological Inflation Attractors}},
	volume = {114},
	year = {2015}}

@article{Terada:2016nqg,
	archiveprefix = {arXiv},
	author = {Terada, Takahiro},
	doi = {10.1016/j.physletb.2016.07.058},
	eprint = {1602.07867},
	journal = {Phys. Lett. B},
	pages = {674--680},
	primaryclass = {hep-th},
	reportnumber = {UT-16-10, APCTP-PRE2016-005},
	title = {{Generalized Pole Inflation: Hilltop, Natural, and Chaotic Inflationary Attractors}},
	volume = {760},
	year = {2016}}

@misc{paper,
author = {Freese, Katherine and Sfakianakis, Evangelos I. and Shams Es Haghi, Barmak},
note = {To appear},
	year = {2026}}

@article{Zhitnitsky:1980tq,
    author = "Zhitnitsky, A. R.",
    title = "{On Possible Suppression of the Axion Hadron Interactions. (In Russian)}",
    journal = "Sov. J. Nucl. Phys.",
    volume = "31",
    pages = "260",
    year = "1980"
}

@article{Dine:1981rt,
    author = "Dine, Michael and Fischler, Willy and Srednicki, Mark",
    title = "{A Simple Solution to the Strong CP Problem with a Harmless Axion}",
    reportNumber = "Print-81-0320 (IAS,PRINCETON)",
    doi = "10.1016/0370-2693(81)90590-6",
    journal = "Phys. Lett. B",
    volume = "104",
    pages = "199--202",
    year = "1981"
}

@article{Jeong:2013xta,
    author = "Jeong, Kwang Sik and Takahashi, Fuminobu",
    title = "{Suppressing Isocurvature Perturbations of QCD Axion Dark Matter}",
    eprint = "1304.8131",
    archivePrefix = "arXiv",
    primaryClass = "hep-ph",
    reportNumber = "TU-935",
    doi = "10.1016/j.physletb.2013.10.061",
    journal = "Phys. Lett. B",
    volume = "727",
    pages = "448--451",
    year = "2013"
}

@article{Higaki:2014ooa,
    author = "Higaki, Tetsutaro and Jeong, Kwang Sik and Takahashi, Fuminobu",
    title = "{Solving the Tension between High-Scale Inflation and Axion Isocurvature Perturbations}",
    eprint = "1403.4186",
    archivePrefix = "arXiv",
    primaryClass = "hep-ph",
    reportNumber = "KEK-TH-1711, DESY-14-033, TU-958, IPMU14-0060",
    doi = "10.1016/j.physletb.2014.05.014",
    journal = "Phys. Lett. B",
    volume = "734",
    pages = "21--26",
    year = "2014"
}

@article{vonHarling:2017yew,
    author = "von Harling, Benedict and Servant, Geraldine",
    title = "{QCD-induced Electroweak Phase Transition}",
    eprint = "1711.11554",
    archivePrefix = "arXiv",
    primaryClass = "hep-ph",
    reportNumber = "DESY-17-056",
    doi = "10.1007/JHEP01(2018)159",
    journal = "JHEP",
    volume = "01",
    pages = "159",
    year = "2018"
}

\onecolumngrid

\vspace{0.5em}
\begin{center}
    \textbf{\large Appendix}
\end{center}
\vspace{0.8em}

\twocolumngrid

{\bf Axion de Sitter fluctuations:} The dynamics can be understood via stochastic inflation. The equation of motion for the coarse-grained axion angle $\theta = a/f_a$ in de Sitter space is:
\begin{equation}
f_a^2 {\partial \over\partial \Delta N}\langle  \theta^2\rangle = {H^2\over 4\pi^2} - \frac{2 m_a^2}{3H^2} f_a^2 \langle  \theta^2\rangle \, .
\end{equation}
For $m_a \gg H$, the field is driven to zero. Once deconfinement occurs and the axion becomes light ($m_a \ll H$), the variance grows linearly with the number of $e$-folds for which the axion is light ($\Delta N$), yielding $\sigma_\theta^2 = (H^2/4\pi^2 f_a^2) \Delta N$.

\medskip

{\bf Equivalence of our parametrization ($p=2$) to $\alpha$-Attractors:} To demonstrate the robustness of our generic $p=2$ parameterization, we map it exactly onto the T-model of $\alpha$-attractor inflation. First, let us derive the field excursion starting from our generic parameterization, $\epsilon(N) = A_\epsilon  / N^2$. From the definition of the first slow-roll parameter $\epsilon\equiv -\dot H/H^2$, we derive the  relation $d\phi/dN = - M_{\rm Pl}\sqrt{2\epsilon(N)}$, which we integrate to find the field profile of Eq.~\eqref{eq:p2phi}.

We will now compare Eq.~\eqref{eq:p2phi} to the explicit dynamics of the T-model, defined by $V(\phi) = V_0 \tanh^2(\phi/\sqrt{6\alpha})$. Due to the flatness of the potential away from its minimum ($|\phi|/\sqrt{\alpha} \gg 1$), we take $V \simeq V_0$ during slow-roll inflation, yielding $H \simeq \sqrt{V_0/3}/M_{\rm Pl}$. The potential derivative is $V_{,\phi} \simeq 4 \sqrt{2/3\alpha} V_0 e^{\sqrt{2/3\alpha} \phi}$. Integrating the $e$-folding number $dN = H dt = (H/\dot{\phi}) d\phi$ yields the explicit field value:
\begin{equation}
    \phi(N) = -\sqrt{\frac{3\alpha}{2}} \ln\left( \frac{8 M_{\rm Pl}^2}{3\alpha} N \right) = -\sqrt{\frac{3\alpha}{2}} \ln N - C \, ,
    \label{eq:phi_exact_Tmodel}
\end{equation}
where $C = \sqrt{3\alpha/2} \ln( 8 M_{\rm Pl}^2 / 3\alpha )$ acts the integration constant. Calculating the first slow-roll parameter explicitly yields $\epsilon(N) \simeq {3\alpha} / ({4 M_{\rm Pl}^2} {N^2})$. 

Comparing this to our parameterization rigorously establishes the  relation $A_\epsilon  \equiv {(3\alpha)}/{(4 M_{\rm Pl}^2)}$. If we substitute this  back into the generic field excursion (Eq.~\eqref{eq:p2phi}), the prefactor of the logarithm becomes $\sqrt{3\alpha/2}$. Thus, our generic parameterization perfectly recovers the slow-roll T-model dynamics.

\medskip

{\bf Slow-Roll analysis of Quartic Hilltop Inflation:}
Let us consider the potential of quartic hilltop inflation
\begin{equation}
\label{eq:hilltoppotential}
    V(\phi) = \Lambda^4 \left( 1 - \frac{\phi^4}{\mu^4} \right) \, ,
\end{equation}
where $\Lambda$ determines the energy scale of inflation and $\mu$ sets the width of the plateau. We assume the field is rolling away from the origin at $\phi = 0$.
This can be made compatible with the mechanism of raising the QCD scale, by altering the coupling term in Eq.~\eqref{eq:coupling} from $(1/ M) \phi G_{\mu\nu}G^{\mu\nu}$ to $(1/ M) (\phi-\phi_{\rm end}) G_{\mu\nu}G^{\mu\nu}$. In the latter case, at the end of inflation $\phi=\phi_{\rm end}$ and we recover the usual gluon kinetic term. We can alternatively shift the field value in the hilltop potential as $V\propto (1-(\phi-\phi_{\rm end})^4/\mu^2 )$, keeping Eq.~\eqref{eq:coupling} intact. Both field redefinitions lead to the same dynamics.
 The  potential of Eq.~\eqref{eq:hilltoppotential} can be a valid approximation only for $|\phi|<\mu$, as it becomes negative for $|\phi|>\mu$ and thus higher order terms are needed to stabilize it. These terms can alter the field value at the end of inflation, which in turn  shifts the field values and / or the numbering of the $e$-folds ($\phi_\infty$ and $N_0$ in Eq.~\eqref{eq:p3phi}). 
In the  plateau approximation, we assume that inflation occurs strictly in the small-field regime, $\phi \ll \mu$, where possible higher order stabilizing terms can be ignored. Therefore, the energy density is dominated by the constant term, $V(\phi) \approx \Lambda^4$, while the dynamics are governed by the first and second slow-roll parameters
(again in the $\phi \ll \mu$ limit)
\begin{align}
    \epsilon &= \frac{M_{\mathrm{Pl}}^2}{2} \left( \frac{V'}{V} \right)^2 \approx 8 M_{\mathrm{Pl}}^2 \frac{\phi^6}{\mu^8} \, , \\[6pt]
    \eta &= M_{\mathrm{Pl}}^2 \frac{V''}{V} \approx -12 M_{\mathrm{Pl}}^2 \frac{\phi^2}{\mu^4} \, .
\end{align}
In the regime $\phi \ll \mu$, we see that the condition $\epsilon \ll |\eta|$ is trivially satisfied, as expected for a plateau model.

Inflation ends  when the first slow-roll parameter reaches unity, $\epsilon(\phi_{\mathrm{end}}) = 1$. Solving for the field value at the end of inflation, using the expression of $\epsilon$ derived in the $\phi\ll \mu$ regime yields
\begin{equation}
    \phi_{\mathrm{end}} = \frac{\mu^{4/3}}{\sqrt{2} M_{\mathrm{Pl}}^{1/3}} \, .
\end{equation}
 Let us note that for the small-field approximation $\phi_{\mathrm{end}} \ll \mu$ to remain globally valid, the theory  requires  $\mu \ll M_{\mathrm{Pl}}$. If inflation  proceeds beyond the plateau approximation, allowing for $\mu>M_{\rm Pl}$, the following analysis needs to be modified~\cite{Dimopoulos:2020kol, German:2020rpn, Adshead:2010mc}.

The number of $e$-folds $N$ between horizon crossing ($\phi_*$) and the end of inflation ($\phi_{\mathrm{end}}$) is given by:
\begin{equation}
    N =   \int_{t_{\rm init}}^{t_{\rm end}} H \, dt
    = \int_{\phi_*}^{\phi_{\mathrm{end}}} {H\over \dot\phi}\, d\phi \approx \frac{\mu^4}{4 M_{\mathrm{Pl}}^2} \int_{\phi_*}^{\phi_{\mathrm{end}}} \phi^{-3} \, d\phi
\end{equation}
Evaluating the integral gives:
\begin{equation}
    N \approx \frac{\mu^4}{8 M_{\mathrm{Pl}}^2} \left( \frac{1}{\phi_*^2} - \frac{1}{\phi_{\mathrm{end}}^2} \right)
\end{equation}
Substituting $\phi_{\mathrm{end}}^2 = \mu^{8/3} / (2 M_{\mathrm{Pl}}^{2/3})$ into the expression, we can solve for the field value at horizon crossing:
\begin{equation}
    \phi_*^2 \approx \frac{\mu^4}{8 M_{\mathrm{Pl}}^2 \left[ N + \frac{1}{4} \left(\frac{\mu}{M_{\mathrm{Pl}}}\right)^{4/3} \right]} \equiv \frac{\mu^4}{8 M_{\mathrm{Pl}}^2 N_{\mathrm{eff}}}
\end{equation}
where $N_{\mathrm{eff}} = N + \frac{1}{4}(\mu/M_{\mathrm{Pl}})^{4/3}$ absorbs the finite correction from the end of inflation. Since $\mu\ll M_{\rm Pl}$, we can  drop this correction and recover $N_{\rm eff}\simeq N$.

Evaluating the slow-roll parameters at horizon crossing $\phi_*$, we find
\begin{align}
    \eta_* &\approx -12 \frac{M_{\mathrm{Pl}}^2}{\mu^4} \left( \frac{\mu^4}{8 M_{\mathrm{Pl}}^2 N_{\mathrm{eff}}} \right) = -\frac{3}{2 N_{\mathrm{eff}}} \simeq -{3\over 2N} \, ,\\
    \nonumber
    \epsilon_* &\approx 8 \frac{M_{\mathrm{Pl}}^2}{\mu^8} \left( \frac{\mu^4}{8 M_{\mathrm{Pl}}^2 N_{\mathrm{eff}}} \right)^3 
    \simeq  \frac{1}{64 N^3} \left(\frac{\mu}{M_{\mathrm{Pl}}}\right)^4\, ,
\end{align}
where we took the limit $\mu\ll M_{\rm Pl}$ in the last step of the two above equations.
The scalar spectral index $n_s$ and the tensor-to-scalar ratio $r$ evaluated at first order in slow-roll are
\begin{align}
    n_s &= 1 + 2\eta_* - 6\epsilon_* \approx 1 - \frac{3}{N_*}  \, ,\\
    r &= 16\epsilon_* \approx \frac{1}{4 N^3_*} \left(\frac{\mu}{M_{\mathrm{Pl}}}\right)^4 \, .
\end{align}
For $N=60$ the tensor to scalar ratio is unobservably small (unless $\mu>M_{\rm Pl}$), $r={\cal O}(10^{-6})\mu^4/M_{\rm Pl}^4\ll 1$, and the spectral index is in strong tension with current CMB measurements, $n_s\simeq 0.95$.

\medskip

{\bf Equivalence of our parametrization ($p=3$) to Quartic Hilltop Inflation:} To demonstrate the practicality of our generic $p=3$ parameterization, we map it exactly onto the Quartic Hilltop model of inflation, which we described above. First, let us derive the field excursion independently from our generic parameterization, $\epsilon(N) = A_\epsilon  / N^3$. Integrating the relation $d\phi/dN = -M_{\rm Pl}\sqrt{2\epsilon(N)}$ , we find the field profile of Eq.~\eqref{eq:p3phi}.
We see that Eq.~\eqref{eq:p3phi} matches the slow-roll solution for plateau quartic hilltop inflation, if we set $\phi_\infty=0$ and $N+N_0=N_{\rm eff}. $

Similarly  comparing the first slow-roll parameter for quartic hilltop inflation to our parameterization  establishes the  relation $A_\epsilon  \equiv {\mu^4}/({64 M_{\rm Pl}^4})$. If we substitute this relation   into Eq.~\eqref{eq:p3phi} with $\phi_{\rm infty}=0$, we recover the exact Quartic Hilltop trajectory. Let us note that $A_\epsilon>1$ requires $\mu>M_{\rm Pl}$. This is allowed within the $p=3$ family, but the connection to hilltop inflation requires going beyond the simple plateau approximation.

\medskip

{\bf Equivalence of our parametrization ($p=3$) to Kinetic Pole  Inflation:}
In pole inflation (often arising in supergravity or string theory compactifications), the inflaton field $\rho$ has a non-canonical kinetic term. The general Lagrangian is given by
\begin{equation}
    \mathcal{L} = -\frac{1}{2} \frac{a}{\rho^q} (\partial \rho)^2 - V(\rho) ~,
\end{equation}
where $q$ is the order of the kinetic pole at $\rho = 0$, and $a$ is a dimensionless constant. 

To ensure that inflation occurs as the field rolls away from the pole, the potential must be regular and possess a local maximum at $\rho = 0$. We can Taylor expand the potential near the pole as
\begin{equation}
    V(\rho) \simeq V_0 \left( 1 - c \rho^n \right) ~,
\end{equation}
where $n$ is the leading non-zero power in the expansion, and $c > 0$.

To connect this framework to the generic parametrization $\epsilon(N) = A_\epsilon / N^p$ with $p=3$, we must first transform the field $\rho$ into a canonically normalized field $\phi$. The relation is given by
\begin{equation}
    d\phi = \frac{\sqrt{a}}{\rho^{q/2}} d\rho ~.
\end{equation}
Assuming $q < 2$, we can integrate this expression starting from the pole:
\begin{equation}
      \rho = \left( \frac{(1 - q/2) \phi}{\sqrt{a}} \right)^{\frac{2}{2-q}} ~.
\end{equation}
Notice that the pole at $\rho \to 0$ maps directly to the origin $\phi \to 0$. Substituting this inverse relation back into the potential yields the effective canonical potential:
\begin{equation}
    V(\phi) = V_0 \left[ 1 - c \left( \frac{1 - q/2}{\sqrt{a}} \right)^{\frac{2n}{2-q}} \phi^{\frac{2n}{2-q}} \right] ~.
\end{equation}
We have already demonstrate that the $p=3$ universality class corresponds exactly to the Quartic Hilltop model $V\propto 1-\phi^4/\mu^4$.
Therefore, to  reproduce the $p=3$ dynamics using a kinetic pole, the exponent of $\phi$ in our derived canonical potential must be equal to 4, leading to
\begin{equation}
   {q = 2 - \frac{n}{2}} \,. \label{eq:polemaster_relation}
\end{equation}
Equation (\ref{eq:polemaster_relation}) provides a direct dictionary for constructing $p=3$ inflation from a kinetic pole. Depending on the geometry of the potential at the maximum, there are two primary physical realizations:

If the potential has a  quadratic maximum at the pole ($n=2$), then $V(\rho) = V_0(1 - c \rho^2)$. According to our relation, this requires a {simple pole of order $q=1$}.
The field transformation becomes $ \rho = {\phi^2}/{4a}$
    and the potential of the canonically normalized field  is $V(\phi) = V_0 \left( 1 - \frac{c}{16 a^2} \phi^4 \right)$.  This precisely matches the Quartic Hilltop model. By setting $\Lambda^4 = V_0$ and the width parameter to $\mu^4 = \frac{16 a^2}{c}$, we can write the generic slow-roll amplitude $A_\epsilon$  in terms of the pole parameters as
  $     A_\epsilon = {a^2}/{4 c M_{\rm Pl}^4} $.

If instead the potential is linear at the pole ($n=1$), meaning $V(\rho) = V_0(1 - c \rho)$, this requires a {fractional pole of order $q=3/2$}.
The field transformation becomes $\rho = {\phi^4}/{256 a^2}$
   and the potential of the canonically normalized field is $V(\phi) = V_0 \left( 1 - \frac{c}{256 a^2} \phi^4 \right)$. Finally, the constant in the slow-roll parametrization becomes 
  $
        A_\epsilon = {4 a^2}/{c M_{\rm Pl}^4}$.

\end{document}